\documentclass[prb,aps,twocolumn,superscriptaddress]{revtex4-1}

\usepackage{color,amsmath,amssymb,amsfonts,graphicx,tabularx}

\usepackage[unicode=true,colorlinks=true]{hyperref}

\hypersetup{linkcolor=blue,citecolor=blue,urlcolor=blue}

\newcommand{\overbar}[1]{\mkern 2.2mu\overline{\mkern-2.2mu#1\mkern-2.2mu}\mkern 2.2mu}

\begin{document}

\title{Fermionic symmetry-protected topological state in strained graphene}

\author{Ying-Hai Wu}
\affiliation{Max-Planck-Institut f{\"u}r Quantenoptik, Hans-Kopfermann-Stra{\ss}e 1, 85748 Garching, Germany}

\author{Tao Shi}
\affiliation{Max-Planck-Institut f{\"u}r Quantenoptik, Hans-Kopfermann-Stra{\ss}e 1, 85748 Garching, Germany}

\author{G. J. Sreejith}
\affiliation{Max-Planck-Institut f{\"u}r Physik komplexer Systeme, 01187 Dresden, Germany}
\affiliation{Indian Institute for Science Education and Research, Pune 411008, India}

\author{Zheng-Xin Liu}
\affiliation{Renmin University of China, Beijing 100876, China}

\begin{abstract}
The low-energy physics of graphene is described by relativistic Dirac fermions with spin and valley degrees of freedom. Mechanical strain can be used to create a pseudo magnetic field pointing to opposite directions in the two valleys. We study interacting electrons in graphene exposed to both an external real magnetic field and a strain-induced pseudo magnetic field. For a certain ratio between these two fields, it is proposed that a fermionic symmetry-protected topological state can be realized. The state is characterized in detail using model wave functions, Chern-Simons field theory, and numerical calculations. Our paper suggests that graphene with artificial gauge fields may host a rich set of topological states.
\end{abstract}

\maketitle

\section{Introduction}

The study of topological phases has been a central topic of condensed matter physics since the observations of the quantum Hall effect \cite{Klitzing1980,Tsui1982}. The discovery of topological insulators signifies that the interplay between topology and symmetry can lead to a variety of exotic phenomena \cite{Hasan2010,QiXL2011-1}. The concept of symmetry-protected topological (SPT) states has been introduced to describe gapped quantum states which are non-trivial only if certain symmetries are enforced. The SPT states are different from topologically ordered states in that they do not exhibit fractionalization or long-range entanglement, while they are distinguished from trivial product states by their non-trivial edge states if the protecting symmetries are not broken explicitly or spontaneously.

For non-interacting fermions, the possible SPT states have been fully classified for various symmetries \cite{Schnyder2008,Kitaev2009}. It is natural to ask what happens if interactions are introduced. On one hand, distinct SPT states in free fermion systems may be adiabatically connected if interactions are allowed \cite{Fidkowski2011}. On the other hand, new SPT states with no free fermion counterparts may emerge because of interactions. The SPT states in one- and two-dimensional interacting bosonic systems can be constructed and classified by the group cohomology theory \cite{ChenX2011,Schuch2011,ChenX2012}, while SPT states beyond group cohomology in three dimensions have been reported \cite{Vishwanath2013}. For two-dimensional systems, the Chern-Simons field theory turns out to be very useful for classifying the SPT states and studying their physical properties \cite{LuYM2012}. As an example of bosonic SPT states, the integer quantum Hall (IQH) states of bosons have been identified in several microscopic models \cite{Senthil2013,FurukawaS2013,WuYH2013,Regnault2013,Sterdyniak2015,HeYC2015,ZengTS2016}. The theory of fermionic SPT states is less complete except for in one dimension \cite{GuZC2014-2,YouYZ2014,Kapustin2015,ChengM2015}. One exotic possibility is that topological orders may be found on the surfaces of three-dimensional SPT states in some ways that are impossible in strictly two-dimensional systems \cite{Bonderson2013,WangC2013,ChenX2014,Metlitski2015}.

In this work, we construct a fermionic SPT state in two dimensions and analyze its properties in detail. This state depends essentially on interactions because its physical responses cannot appear in the same setup without interactions. The occurrence of this state requires two types of fermions coupled to two different magnetic fields. One possible platform fulfilling such a condition is strained graphene. The band structure of graphene contains two inequvialent valleys ${\mathbf K}^{\pm}$ in the Brillouin zone where the physics is described by Dirac fermions with linear dispersion. By applying suitably designed strain, a pseudo magnetic field pointing to opposite directions in the two valleys can be generated \cite{Guinea2010,Levy2010,Ghaemi2012,Abanin2012-2,Gradinar2013,Roy2013,Verbiest2015,LiSY2015,Georgi2017}. If an external magnetic field is also applied, the electrons in the two valleys would experience different total magnetic fields. This paper is organized as follows: the model of our interest is defined in Sec. \ref{Model}, the fermionic SPT state is characterized using wave function and field theory in Sec. \ref{State}, the relevance of this state in strained graphene is corroborated by numerical calculations in Sec. \ref{Result}, and we conclude in Sec. \ref{Conclusion}.

\section{Model}
\label{Model}

\begin{figure}
\includegraphics[width=0.45\textwidth]{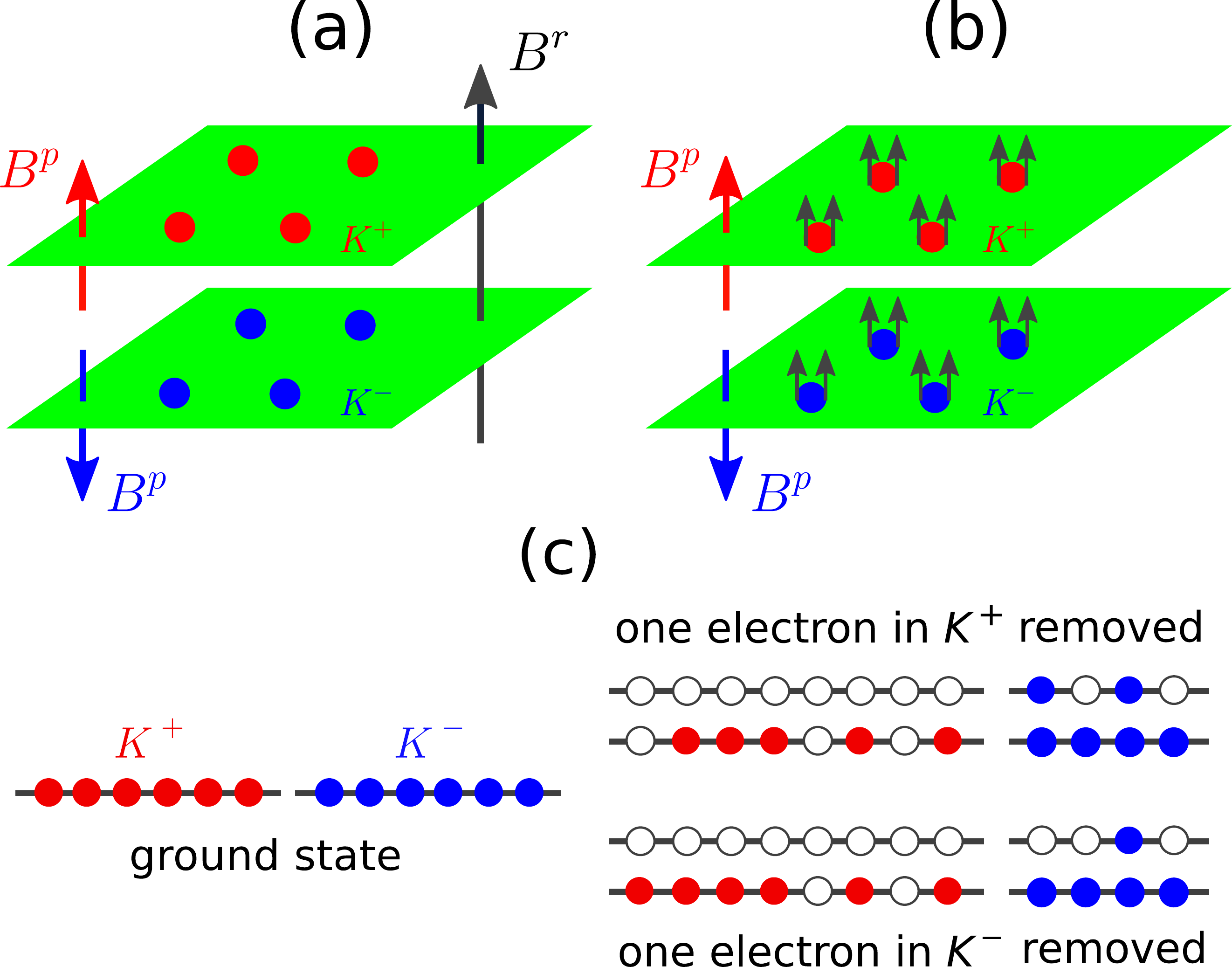}
\caption{(a) The model of graphene with a external real magnetic field $B^{r}$ and a strain induced pseudo magnetic field $B^{p}$. (b) The electrons in the two valleys each absorb two fluxes and become composite fermions moving in opposite effective magnetic fields. (c) The consequences of removing one electron from the ground state.}
\label{Figure1}
\end{figure}

We define the graphene sheet as the $xy$ plane, the unit vector perpendicular to the $xy$ plane as ${\widehat e}_{z}$, and the electron charge as $-e$. The system experience both a real magnetic field and a strain induced pseudo magnetic field as shown in Fig.~\ref{Figure1} (a). The real (pseudo) magnetic field is denoted as $B^{r}$ ($B^{p}$) and the corresponding vector potential is ${\mathbf A}^{r}$ (${\mathbf A}^{p}$). It is assumed that $B^{r}>B^{p}$ and the total magnetic fields $B^{\pm}=B^{r}{\pm}B^{p}$ in the two valleys point to the $-{\widehat e}_{z}$ direction. The single-particle Hamiltonians in the two valleys are
\begin{eqnarray}
{\mathcal H}_{{\mathbf K}^{\pm}} &=& v_{\rm F}
\left[
\begin{array}{cc}
0 & \pi^{\pm}_{x}{\pm}i\pi^{\pm}_{y} \\
\pi^{\pm}_{x}{\mp}i\pi^{\pm}_{y} & 0 
\end{array} 
\right]
\end{eqnarray}
where $v_{\rm F}$ is the Fermi velocity and $\pi^{\pm}_{i}=p_{i}-e(A^{r}_{i}{\pm}A^{p}_{i})/c$ are the canonical momentum operators. The solutions in each valley contain a set of zero energy states, whose spatial components are the same as the non-relativistic lowest Landau level (LL) wave functions.

For an infinite disk with symmetric gauge, the non-relativistic lowest LL wave functions have the simple forms $\psi_{m}(x,y) \sim z^{m}\exp[-|z|^2/(4\ell^2_{B})]$, where $\ell_{B}=\sqrt{hc/(eB)}$ is the magnetic length, $z=(x+iy)$ is the complex coordinate, and $m$ is the angular momentum. If the direction of the magnetic field is reversed, the solutions are complex conjugates $\psi^{*}_{m}(x,y)$. The electrons interact with each other via the Coulomb potential $e^2/(\varepsilon|{\mathbf r}_{j}-{\mathbf r}_{k}|)$ ($\varepsilon$ is the dielectric constant of the system). The magnetic field is taken to be sufficiently strong so the electrons are confined to the zero energy states. The many-body problem can be studied on compact surfaces such as sphere and torus \cite{Haldane1983,Yoshioka1983} to avoid edge effects (see Appendix for some technical details). The low-energy states are assumed to be spin-polarized due to the Zeeman splitting and/or quantum Hall ferromagnetism \cite{NomuraK2006,Goerbig2006,Alicea2006} (this will be corroborated by numerical results later). The numbers of electrons in the two valleys are denoted as $N^{\pm}_{e}$. The system respects a $U(1)_{r}{\times}U(1)_{p}$ symmetry, where $U(1)_{r}$ [$U(1)_{p}$] corresponds to the conservation of $N^{+}_{e}+N^{-}_{e}$ ($N^{+}_{e}-N^{-}_{e}$). The numbers of magnetic fluxes through the surface of sphere or torus in the two valleys are denoted as $N^{\pm}_{\phi}$. The magnetic length $\ell^{+}_{B}=\sqrt{hc/(eB^{+})}$ associated with $B^{+}$ is used as the length scale and $e^2/(\varepsilon\ell^{+}_{B})$ is used as the energy scale. For electrons on the sphere, an angular momentum quantum number $L$ labels the many-body eigenstates. For electrons on the torus, special momentum quantum number $Y$ labels the many-body eigenstates. 

\section{Topological State}
\label{State}

\subsection{Wave Function}

To construct a many-body state, we start from a non-interacting system with $N^{+}_{e}=N^{-}_{e}$ and choose $B^{p}$ appropriately such that the two valleys form two decoupled IQH states at filling factor $1$ with opposite chiralities ($B^{r}$ is zero at this stage). This system has the wave function
\begin{eqnarray}
\prod_{j<k} (z^{+}_{j}-z^{+}_{k}) \prod_{j<k} (z^{-}_{j}-z^{-}_{k})^*
\end{eqnarray}
where the superscripts $\pm$ are used to label the two valleys. As we turn on the real magnetic field and the interaction between electrons, the composite fermion theory \cite{Jain1989-1} suggests that the interaction energy can be efficiently minimized if the electrons each absorb two magnetic fluxes as shown in Fig.~\ref{Figure1} (b). This transformation is implemented by the factor $\prod_{j<k} (z_{j}-z_{k})^2$, where the coordinates without superscripts are for {\em all} electrons. If we choose $B^{r}=4B^{p}$, the real magnetic field is completely absorbed by the electrons and the resulting composite fermions form two decoupled IQH states. The wave function for this system is
\begin{eqnarray}
{\mathcal P}_{\rm LLL} \prod_{j<k} (z^{+}_{j}-z^{+}_{k}) \prod_{j<k} (z^{-}_{j}-z^{-}_{k})^* \prod_{j<k} (z_{j}-z_{k})^2
\label{ManyWave}
\end{eqnarray}
where ${\mathcal P}_{\rm LLL}$ projects wave functions to the lowest LLs (i.e. the zero energy single-particle states in the two valleys). The filling factors in the two valleys are $\nu^{+}=1/5$ and $\nu^{-}=1/3$ respectively.

The mapping to composite fermions also helps us to understand the excitations because the composite fermions can be taken as non-interacting objects and form their effective LLs. For the ground state, the two types of composite fermions completely fill their respective lowest effective LLs. A neutral excitation is created if we promote one composite fermion from its lowest LL to second LL. There are four types of charged minimal excitations: one type I (II) quasihole is present if the composite fermions occupy all orbitals of the lowest LLs except for one orbital in the ${\mathbf K}^{+}$ (${\mathbf K}^{-}$) valley; one type I (II) quasiparticle is present if the composite fermions occupy all orbitals of the lowest LLs and one orbital of the second LL in the ${\mathbf K}^{+}$ (${\mathbf K}^{-}$) valley. The local charges of the type I (II) quasihole/quasiparticle with respect to ${\mathbf A}^{r}$ are denoted as ${\pm}Q^{r}_{\rm I}$ (${\pm}Q^{r}_{\rm II}$). The local charges with respect to ${\mathbf A}^{p}$ are denoted similarly with the superscript $r$ replaced by $p$. Let us consider what happens if one electron is removed from the ground state [Fig. \ref{Figure1} (c)]. The effective magnetic fluxes for the composite fermions in the ${\mathbf K}^{+}$ (${\mathbf K}^{-}$) valley increase (decrease) by two units. If the electron is from the ${\mathbf K}^{+}$ valley, the real and pseudo charges increase by $e$ and three type I quasiholes and two type II quasiparticles are created. If the electron is from the ${\mathbf K}^{-}$ valley, the real (pseudo) charge increases (decreases) by $e$ and two type I quasiholes and one type II quasiparticles are created. This analysis yields the equations
\begin{eqnarray}
&& 3Q^{r}_{\rm I}-2Q^{r}_{\rm II}=e \;\;\; 2Q^{r}_{\rm I}-Q^{r}_{\rm II}=e \nonumber \\
&& 3Q^{p}_{\rm I}-2Q^{p}_{\rm II}=e \;\;\; 2Q^{p}_{\rm I}-Q^{p}_{\rm II}=-e 
\end{eqnarray}
so we have $Q^{r}_{\rm I}=e$, $Q^{r}_{\rm II}=e$, $Q^{p}_{\rm I}=-3e$, and $Q^{p}_{\rm II}=-5e$, which demonstrates that the charges of electron do not fractionalize in our system.

The Hall conductances can be obtained using the Laughlin flux insertion argument \cite{Laughlin1983}. We place the system on a disk and insert one flux at the center. If the inserted flux is for the real magnetic field (which increases the flux values in both valleys), one type I quasihole and one type II quasiparticle are created at the center. The change of real charge at the center is $Q^{r}_{\rm I}-Q^{r}_{\rm II}=0$ so the real Hall conductance $\sigma^{r}_{xy}=0$. If the inserted flux is for the pseudo magnetic field [which increases (decreases) the flux value in the ${\mathbf K}^{+}$ (${\mathbf K}^{-}$) valley], one type I quasihole and one type II quasihole are created at the center. The change of pseudo charge at the center is $Q^{p}_{\rm I}+Q^{p}_{\rm II}=-8e$ so the pseudo Hall conductance $\sigma^{p}_{xy}=-8e^2/h$. To measure the change of real (pseudo) charge due to the insertion of a pseudo (real) flux, we define a mutual Hall conductance $\sigma^{rp}_{xy}$ whose value turns out to be $2e^2/h$. It should be emphasized that these three Hall conductance values cannot appear simultaneously in a system of non-interacting two-component electrons, where two decoupled IQH states with the same or opposite chiralities at any filling factors can in principle be realized using real plus pseudo magnetic fields.

The system is expected to possess two counterpropagating edge modes corresponding to the edge modes of the IQH states of composite fermions. These edge modes will not be gapped out if there is no tunneling between the valleys. Another possibility is that the composite fermion edge states will remain gapless so long as an emergent time-reversal symmetry of composite fermions is preserved. This is intuitively plausible because the $\nu={\pm}1$ IQH states can be viewed as the simplest two-dimensional topological insulator. This will be formulated in a more precise way using effective field theory below. However, if electrons can tunnel between the valleys, $N^{\pm}_{e}$, $\sigma^{p}_{xy}$, and $\sigma^{rp}_{xy}$ are no longer well-defined.

\subsection{Field Theory}

The wave function Eq. \ref{ManyWave} can be described by the Chern-Simons theory with Lagrangian density \cite{WenXG1992-1}
\begin{eqnarray}
{\mathcal L}_{1} = \frac{1}{4\pi\hbar} \epsilon^{\lambda\mu\nu} K_{IJ} a_{I\lambda} \partial_{\mu} a_{J\nu} - j_{I\lambda} a_{I\lambda}
\end{eqnarray}
where $a_{I\lambda}$ ($I=1,2$, $\lambda=0,x,y$) are internal gauge fields, $j_{I\lambda}$ is the excitation current, and
\begin{eqnarray}
K = \left(
\begin{array}{cc}
3 & 2 \\
2 & 1
\end{array}
\right)
\label{Kmatrix}
\end{eqnarray}
One important signature of topological phases is the number of degenerate ground states on torus. The existence of multiple degenerate ground states implies the presence of fractionalization. For the Chern-Simons Lagrangian ${\mathcal L}_{1}$, the ground state degeneracy on torus is $|\det{K}|=1$, which is consistent with our conclusion that there are no fractionally charged excitations.

An excitation in the Chern-Simons field theory can be labeled by an integer vector ${\mathbf l}$, which is $({\pm}1,0)^T$ for type I quasiparticle/quasihole and $(0,{\mp}1)^T$ for type II quasiparticle/quasihole. The statistical angle of an excitation labeled by ${\mathbf l}$ with itself is $\theta=\pi{\bf l}^T K^{-1} {\bf l}$, so all the minimal excitations have fermionic self braid statistics. The statistical angle of two excitations labeled by ${\mathbf l_1}$ and ${\mathbf l_2}$ is $\theta_{12} = 2\pi {\bf l_1}^T K^{-1} {\bf l_2}$, so one type I quasihole/quasiparticle and one type II quasihole/quasiparticle have bosonic mutual braid statistics. If two probing $U(1)$ gauge fields $\overbar{A^{r}_{\mu}}$ and $\overbar{A^{p}_{\mu}}$ coupled to the excitation current are introduced, we need to add an extra term
\begin{eqnarray}
{\mathcal L}_{2} = \frac{e}{2\pi\hbar} \epsilon^{\lambda\mu\nu} ( t^{r}_I \overbar{A^{r}_{\lambda}} \partial_{\mu} a_{I\nu} + t^{p}_I \overbar{A^{p}_{\lambda}} \partial_{\mu} a_{I\nu})
\end{eqnarray}
to ${\mathcal L}_{1}$, where the charge vectors ${\mathbf t}^{r}=(1,1)^T$ and ${\mathbf t}^{p}=(1,-1)^T$. The $U(1)$ charges of an excitation labeled by ${\mathbf l}$ are $-e[{\bf t}^{r}]^T K^{-1} {\mathbf l}$ and $-e[{\mathbf t}^{p}]^T K^{-1} {\mathbf l}$, which yield the same results for the minimal excitations as our previous analysis. The Hall conductance with respect to the real gauge field is $\sigma^{r}_{xy} = e^2 [{\mathbf t}^{r}]^{T}K^{-1}{\mathbf t}^{r}/h=0$, the one with respect to the pseudo gauge field is $\sigma^{p}_{xy} = e^2 [{\mathbf t}^{p}]^{T}K^{-1}{\mathbf t}^{p}/h = -8e^2/h$, and the mutual Hall conductance is $\sigma^{rp}_{xy} = e^2 [{\mathbf t}^{r}]^{T}K^{-1}{\mathbf t}^{p}/h = 2e^2/h$, which reproduce the results derived using the Laughlin argument.

For a system described by ${\mathcal L}_{1}+{\mathcal L}_{2}$ in the bulk, it has gapless edge states on an open mainfold as captured by 
\begin{eqnarray}
{\mathcal L}_{\rm edge} = \frac{1}{4\pi\hbar} \left( K_{IJ} \partial_{0} \phi_{I} \partial_{x} \phi_{J} - V_{IJ} \partial_{x} \phi_{I} \partial_{x} \phi_{J} \right)
\end{eqnarray}
where $\phi_{I}$ are chiral boson fields that satisfy the Kac-Moody algebra
\begin{eqnarray}
\left[ \partial_{x} \phi_{I}(x) , \partial_{y} \phi_{J}(y) \right] = 2{\pi}i K^{-1}_{IJ} \partial_{x}\delta(x-y)
\end{eqnarray}
and $V_{IJ}$ depends on the microscopic details at the edge \cite{WenXG1995}. The electron annihilation operators for the two valleys are
\begin{eqnarray}
C_{1} = e^{-i(3\phi_1+2\phi_2)} \;\;\; C_{2} = e^{-i(2\phi_1+\phi_2)} 
\end{eqnarray}
The number of edge modes is the dimension of $K$ and their chiralities are given by the sign of the eigenvalues of $K$. This means that our system has two counterpropagating edge modes and is consistent with our previous analysis based on wave functions. These edge modes can be gapped out by some perturbations, but one can rule out such perturbations by imposing certain symmetries on the system \cite{LuYM2012}. If the system has a symmetry group $G$ and is acted upon by an element $g$ of $G$, the $K$ matrix transforms as
\begin{eqnarray}
K \rightarrow W^{T}_{g} K W_{g} = s_{g} K
\label{KmatrixTransform}
\end{eqnarray}
and the gauge fields $\phi$ transform as
\begin{eqnarray}
\phi \rightarrow W^{-1}_{g} \phi + \delta_{g} \phi
\label{FieldTransform}
\end{eqnarray}
where $W_{g}$ is an integer matrix, $s_{g}$ is $1$ for unitary symmetry and $-1$ for anti-unitary symmetry, and $\delta_{g}\phi$ is a constant. 

If we choose the symmetries to be $U(1)_{r}{\times}U(1)_{p}$, the elements of the symmetry groups can be labeled by $\theta_{r},\theta_{p}\in[0,2\pi)$ and the symmetry operators are
\begin{eqnarray}
{\mathcal U}^{r}_{\theta_{r}}=\exp(-i{\theta_{r}} \sum_{I} t^{r}_{I} C^{\dagger}_{I} C_{I})
\end{eqnarray}
and
\begin{eqnarray}
{\mathcal U}^{p}_{\theta_{p}}=\exp(-i{\theta_{p}} \sum_{I} t^{p}_{I} C^{\dagger}_{I} C_{I})
\end{eqnarray}
The quantities $W_{g}$ and $\delta_{g}\phi$ in Eqs. \ref{KmatrixTransform} and \ref{FieldTransform} can be written as
\begin{eqnarray}
W^{r}_{\theta_{r}} = \left(
\begin{array}{cc}
1 & 0 \\
0 & 1
\end{array}
\right) \;\;\; \delta^{r}_{\theta_{r}} \phi = \theta_{r} \sum_{I} t^{r}_{I} K^{-1}_{IJ}
\end{eqnarray}
for the $U(1)_{r}$ group and
\begin{eqnarray}
W^{p}_{\theta_{p}} = \left(
\begin{array}{cc}
1 & 0 \\
0 & 1
\end{array}
\right) \;\;\; \delta^{p}_{\theta_{p}} \phi = \theta_{p} \sum_{I} t^{p}_{I} K^{-1}_{IJ}
\end{eqnarray}
for the $U(1)_{p}$ group. The edge modes can also be protected by another combination of symmetries. To understand how this works, we convert the $K$ matrix to the diagonal form ${\widetilde K}={\rm Diag}(1,-1)=X^{T}KX$ using
\begin{eqnarray}
X=\left(
\begin{array}{cc}
0 & -1 \\
1 & 2
\end{array}
\right)
\end{eqnarray}
The physics of our system would remain the same if the fields $a_{I\lambda},\phi_{I}$ and the vectors ${\mathbf l}, {\mathbf t}$ are also transformed properly (the transformed ones will be denoted by symbols with a tilde). In the transformed basis, the system is a quantum spin Hall insulator with edge modes that are protected by particle number conservation and time-reversal symmetry. Its symmetry group can be written as $G^{-}_{-}[U(1)_{r},{\mathcal T}]$, where ${\mathcal T}$ is an emergent time-reversal symmetry, the subscript $-$ means ${\mathcal T}^{2}=P_{f}$ (the fermion parity operator), and the superscript $-$ means ${\mathcal U}^{r}_{\theta}{\mathcal T} = {\mathcal T} {\mathcal U}^{r}_{-\theta}$. The time-reversal symmetry operates on ${\widetilde K}$ and ${\widetilde\phi}_{I}$ as
\begin{eqnarray}
&& {\mathcal T} {\widetilde K} {\mathcal T}^{-1} = \sigma_{x} {\widetilde K} \sigma_{x} = -{\widetilde K} \nonumber \\
&& {\mathcal T} {\widetilde\phi}_{1} {\mathcal T}^{-1} = {\widetilde\phi}_{2} \nonumber \\
&& {\mathcal T} {\widetilde\phi}_{2} {\mathcal T}^{-1} = {\widetilde\phi}_{1} + \pi
\end{eqnarray}
As one goes back to the original basis, an emergent time-reversal symmetry can be defined and it operates on $K$ and $\phi_{I}$ as
\begin{eqnarray}
&& {\mathcal T} K {\mathcal T}^{-1} = W^{T}_{\mathcal T} K W_{\mathcal T} = -K \nonumber \\
&& {\mathcal T} \phi_1 {\mathcal T}^{-1} = -2\phi_1 - \phi_2 + \pi \nonumber \\
&& {\mathcal T} \phi_2 {\mathcal T}^{-1} = 3\phi_1 + 2\phi_2
\end{eqnarray}
The quantities $W_{g}$ and $\delta_{g}\phi$ in Eqs. \ref{KmatrixTransform} and \ref{FieldTransform} can be written as
\begin{eqnarray}
W_{\mathcal T} = \left( 
\begin{array}{cc}
-2 & -1 \\
3  & 2
\end{array}
\right) \;\;\; \delta_{\mathcal T} \phi = \left(
\begin{array}{c}
\pi \\
0
\end{array}
\right)
\end{eqnarray}

\section{Numerical Results}
\label{Result}

\begin{figure}
\includegraphics[width=0.48\textwidth]{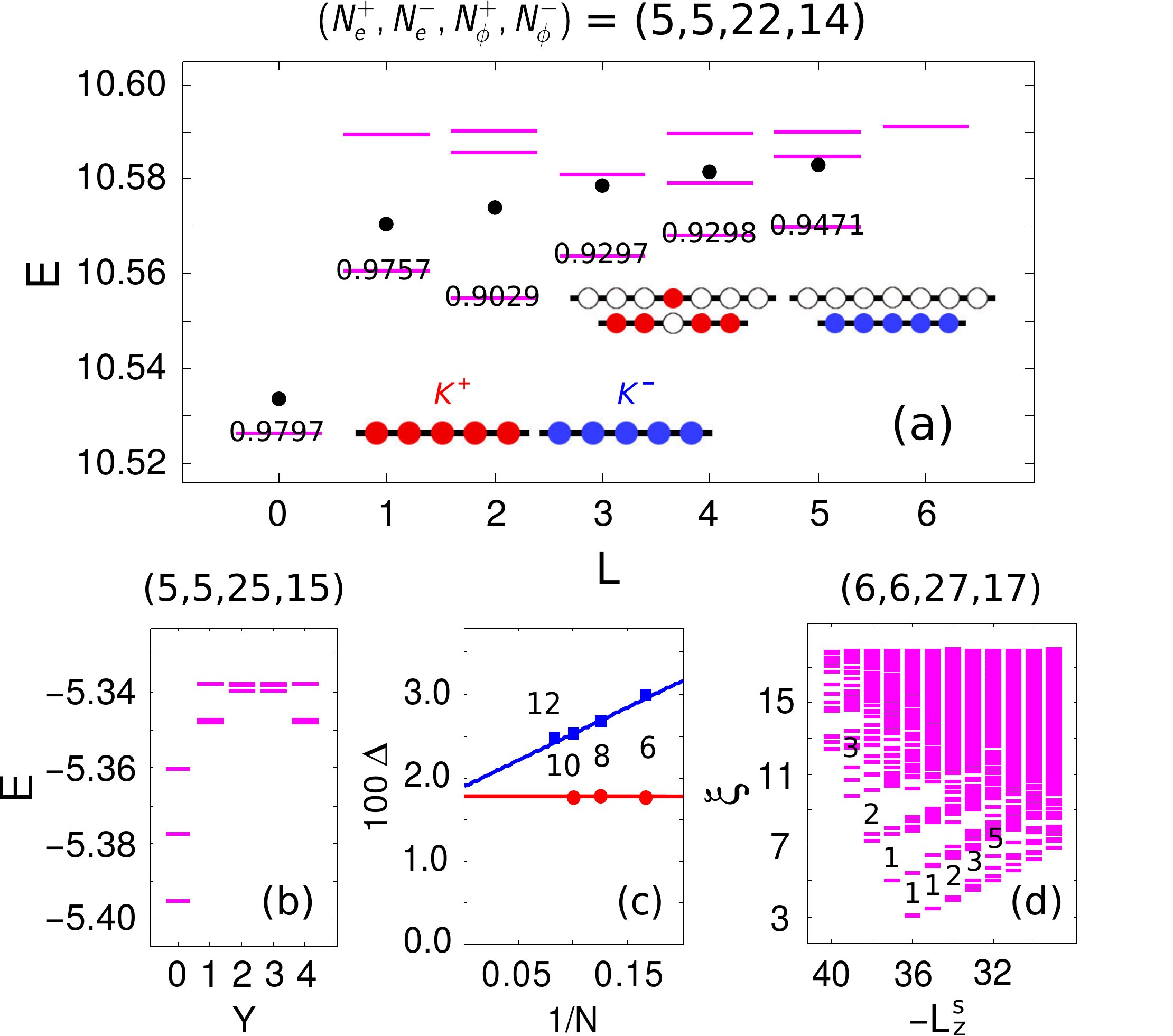}
\caption{(a) Energy spectrum of the Coulomb Hamiltonian on the sphere. The lines and black dots represent the exact eigenstates and the trial states respectively. The overlap between an exact eigenstate and its trial state is shown on the line. The insets give the composite fermion configurations for the low-lying states. (b) Energy spectrum of the Coulomb Hamiltonian on the torus. There is a unique ground state. (c) Energy gap values on the sphere (blue squares, rescaled by the factor $\nu^{+}N^{+}_{\phi}/N^{+}_{e}$ \cite{Morf1986}) and the torus (red dots). The numbers inside the panel are the total numbers of electrons $N=N^{+}_{e}+N^{-}_{e}$ for the data points. The lines are linear fit to the data points. (d) Entanglement spectrum of the Coulomb ground state on the sphere. The numbers of electrons in the southern hemisphere are $N^{S+}_{e}=N^{S-}_{e}=3$. The counting of levels is indicated in the panel. The system parameters of a panel are given as $(N^{+}_{e},N^{-}_{e},N^{+}_{\phi},N^{-}_{\phi})$ on its top.}
\label{Figure2}
\end{figure}

\begin{figure}
\includegraphics[width=0.48\textwidth]{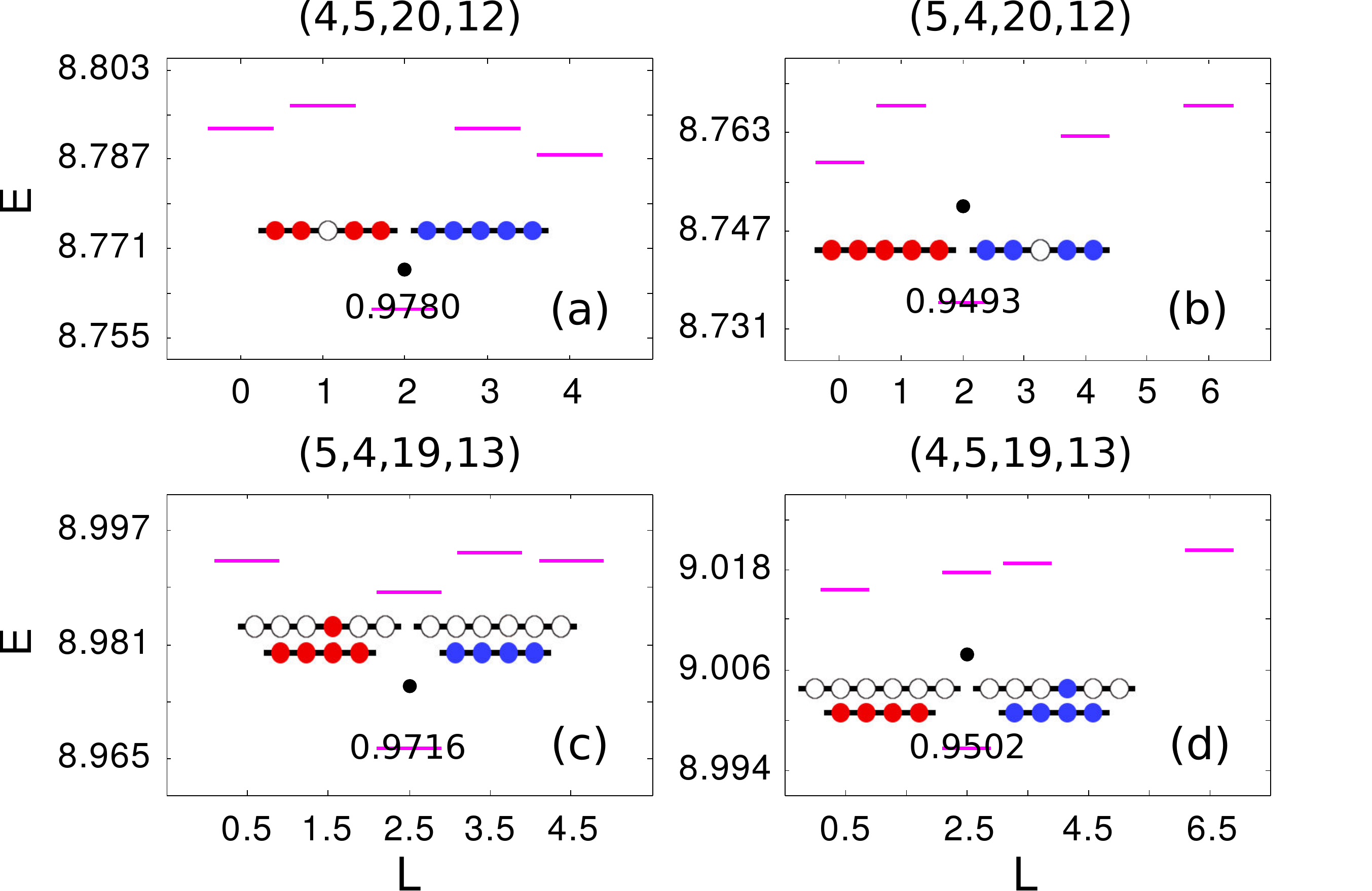}
\caption{Energy spectra of the Coulomb Hamiltonian on the sphere for the cases with (a) one type I quasihole; (b) one type II quasihole; (c) one type I quasiparticle; (d) one type II quasiparticle. The lines, dots, numbers, and insets are defined as in Fig. \ref{Figure1} (a). The system parameters of a panel are given as $(N^{+}_{e},N^{-}_{e},N^{+}_{\phi},N^{-}_{\phi})$ on its top.}
\label{Figure3}
\end{figure}

\begin{figure}
\includegraphics[width=0.45\textwidth]{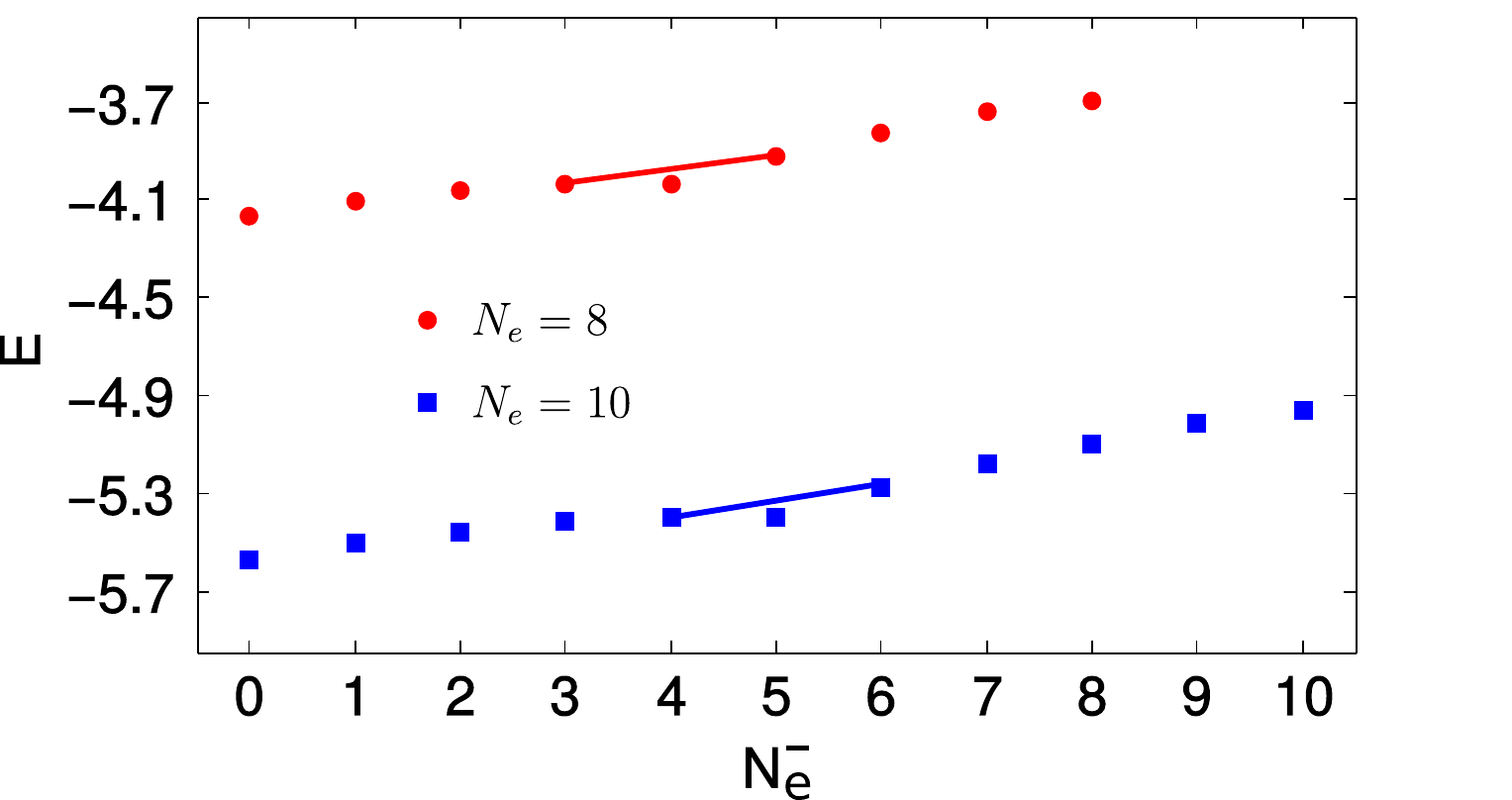}
\caption{The ground state energy on torus versus $N^{-}_{e}$. The total number of electrons is $8$ for the red dots and $10$ for the blue squares.}
\label{Figure4}
\end{figure}

Exact diagonalization can be used to check the validity of our previous analysis. Let us consider the cases where the electrons are spin-polarized and $N^{+}_{e}=N^{-}_{e}$. For electrons on the sphere, the relation between $N^{\pm}_{e}$ and $N^{\pm}_{\phi}$ is $N^{\pm}_{\phi}=N^{\pm}_{e}/\nu^{\pm}-{\mathcal S}^{\pm}$, where ${\mathcal S}^{+}=3$ and ${\mathcal S}^{-}=1$ are the shift quantum numbers. There is no shift quantum number on the torus. Fig. \ref{Figure2} shows the energy spectra of the $N^{+}_{e}=N^{-}_{e}=5$ system on sphere and torus. The existence of a unique ground state on torus is confirmed in all the systems that can be studied. The energy gap $\Delta$ (the energy difference between the first excited state and the ground state) is found to be ${\approx}0.018e^2/(\varepsilon\ell^{+}_{B})$ in the thermodynamic limit based on the finite size scaling analysis in Fig. \ref{Figure2} (c). For $\varepsilon=3$ and $B^{r}=20$ T, the numerical value is about $17$ K. Based on previous experiences \cite{DuRR1993,Dean2008,Ghahari2011}, we expect that disorder and other imperfections in experimental systems would reduce the actual value to $30\%{\sim}50\%$ of the ideal value.

The spherical version of Eq. \ref{ManyWave} can be constructed explicitly and its high overlap with the exact ground state ($0.9797$ for $N^{+}_{e}=N^{-}_{e}=5$ and similar values for smaller systems) corroborates the accuracy of our ansatz. Besides the ground state, there are several low-energy neutral excitations in Fig. \ref{Figure2} (a), which can be modeled by exciting one composite fermion in the ${\mathbf K}^{+}$ valley to an originally unoccupied state. This is an interesting feature that distinguishes our state from fractional quantum Hall states in graphene, for which composite fermions in both valleys contribute to the low-energy neutral excitations \cite{Balram2015-1}. Fig. \ref{Figure3} shows the energy spectra of systems that contain one charged minimal excitation, where the trial wave functions also provide excellent approximations of the exact eigenstates. 

The edge states can be seen from entanglement spectrum \cite{LiH2008}. For this calculation, the sphere is cut along its equator such that two hemispheres are separated by a virtual edge \cite{Dubail2012-1,Sterdyniak2012,Rodriguez2012}. The ground state $|\Psi\rangle$ is decomposed as $|\Psi\rangle = \sum_{ij} F_{ij} |\Psi^{S}_{i}\rangle \otimes |\Psi^{R}_{j}\rangle$, where $S$ ($R$) is the southern hemisphere and its basis states are $|\Psi^{S}_{i}\rangle$ ($|\Psi^{R}_{j}\rangle$). The Schmidt decomposition of $F_{ij}$ gives $|\Psi\rangle = \sum_{\mu} e^{-\xi_{\mu}/2} |\Psi^{S}_{\mu}\rangle \otimes |\Psi^{R}_{\mu}\rangle$ and the $\xi_{\mu}$ levels comprise the entanglement spectrum. The entanglement levels can be labeled by the good quantum numbers $N^{S\pm}_{e}$ (the numbers of electrons) and $L^{S}_{z}$ (the $z$-component angular momentum) of the southern hemisphere. Fig. \ref{Figure2} (d) presents the entanglement spectrum of the $N^{+}_{e}=N^{-}_{e}=6$ system in the $N^{S+}_{e}=N^{S-}_{e}=3$ subspace with the levels organized according to $L^{S}_{z}$. There is a foward-moving branch with counting $1,1,2,3,5$ and a backward-moving branch with counting $1,1,2,3$, which are consistent with the presence of two counterpropagating edge modes each described by a single boson field.

The assumption of spin polarization can also be tested numerically. To this end, we denote the numbers of electrons with spin-valley indices as $N^{\uparrow+}_{e},N^{\downarrow+}_{e},N^{\uparrow-}_{e},N^{\downarrow-}_{e}$. The electrons in the same valley with different spin experience the same magnetic field. For the two valleys, we define total spin operators ${\mathbf S}^2_{\pm}$ and $z$-component spin opertors ${\widehat S}^{z}_{\pm}$. Because the Hamiltonian is $SU(2)$ symmetric in the spin space, the energy eigenstates are also eigenstates of ${\mathbf S}^2_{\pm}$ [with eigenvalues $S_{\pm}(S_{\pm}+1)$] and ${\widehat S}^{z}_{\pm}$ [with eigenvalues $S^{z}_{\pm}=(N^{\uparrow\pm}_{e}-N^{\downarrow\pm}_{e})/2$]. To check for all possible spin polarizations, we should study the subspace with minimal $z$-component spin values $S^{z}_{\pm}$. The ground states for the two systems with $N^{+}_{e}=N^{-}_{e}=3$ and $N^{+}_{e}=N^{-}_{e}=4$ are found to be spin-polarized.

For a fixed $N_{e}=N^{+}_{e}+N^{-}_{e}$, there are many possible choices of $N^{\pm}_{e}$ but only the special cases with $N^{+}_{e}=N^{-}_{e}$ were studied above. It is useful to compare the ground state energy $E_{N^{-}_{e}}$ at different $N^{-}_{e}$. Fig. \ref{Figure4} shows that the lowest one appears at $N^{-}_{e}=0$ in the $N_{e}=8$ and $10$ systems. If a pseudo Zeeman term $\alpha(N^{+}_{e}-N^{-}_{e})$ is added to the Hamiltonian \cite{Georgi2017}, $E_{N^{-}_{e}}$ increases (decreases) for $N^{-}_{e}<N_{e}/2$ ($N^{-}_{e}>N_{e}/2$) and $E_{N_{e}/2}$ will become the global minimum in a certain range of $\alpha$ (estimated to be $0.0344{\lesssim\alpha\lesssim}0.0860$). This can be seen easily from the spectra because the lines connecting all the pairs of data points at $N^{-}_{e}$ and $N_{e}-N^{-}_{e}$ lie above $E_{N_{e}/2}$ (one line is shown explicitly for each case).

The $K$ matrix in Eq. \ref{Kmatrix} suggests another possible trial wave function
\begin{eqnarray}
\prod_{j<k} (z^{+}_{j}-z^{+}_k)^3 \prod_{j<k} (z^{-}_{j}-z^{-}_k) \prod_{j,k} (z^{+}_{j}-z^{-}_k)^2
\label{WaveFunc312}
\end{eqnarray}
for the ground state. It is the zero energy ground state of the Hamiltonian $g_{1} \sum_{j<k} \nabla^2 \delta^{(2)}(\mathbf{r}^{+}_{j} - \mathbf{r}^{+}_{k}) + g_{2} \sum_{j<k} \delta^{(2)}(\mathbf{r}^{+}_{j} - \mathbf{r}^{-}_{k}) + g_{3} \sum_{j<k} \nabla^2 \delta^{(2)}(\mathbf{r}^{+}_{j} - \mathbf{r}^{-}_{k})$, where the positive coefficients $g_{1,2,3}$ determine the interaction strength \cite{Trugman1985}. However, if there is no interaction within the ${\mathbf K}^{-}$ valley but repulsion between the two valleys as required by this parent Hamiltonian, the system would be unstable to phase separation. We have computed the overlap between Eq. \ref{WaveFunc312} and the exact ground state but find that it is much worse than Eq. \ref{ManyWave} ($0.6972$ for the $N^{+}_{e}=N^{-}_{e}=5$ system on sphere).

\section{Conclusion}
\label{Conclusion}

In conclusion, we have studied the properties of a fermionic SPT state in detail and demonstrated that it can be realized in strained graphene. The combination of real and pseudo magnetic fields proposed here allows us to explore a broad range of gauge field configurations for multi-component electrons in graphene. We expect that many other quantum phases in such systems will be revealed. It would also very interesting if one can find some methods to engineer non-Abelian gauge fields and study the quantum phases in such systems.

\section*{Acknowledgement}

Exact diagonalization calculations are performed using the DiagHam package for which we are grateful to all the authors. YHW and TS were supported by the DFG within the Cluster of Excellence NIM. ZXL was supported by the Research Funds of Remin University of China (No. 15XNFL19), NSF of China (No. 11574392), and the Major State Research Development Program of China (2016YFA0300500).

\appendix*

\begin{widetext}

\section{Hamiltonian Matrix Elements}

The two valleys are labeled using $\sigma,\tau=\pm$ and the creation (annihilation) operator for the single-particle state with quantum number $m$ is denoted as $C^\dagger_{\sigma,m}$ ($C_{\sigma,m}$). The electrons interact via the Coulomb potential $V({\mathbf r}_1-{\mathbf r}_2)=e^2/(\varepsilon|{\mathbf r}_1-{\mathbf r}_2|)$. The second quantized form of the many-body Hamiltonian is
\begin{eqnarray}
\frac{1}{2} \sum_{\sigma\tau} \sum_{\{m_{i}\}} F^{\sigma\tau\tau\sigma}_{m_{1}m_{2}m_{4}m_{3}} C^\dagger_{\sigma,m_{1}} C^\dagger_{\tau,m_{2}} C_{\tau,m_{4}} C_{\sigma,m_{3}}
\label{ManyHamilton}
\end{eqnarray}

\subsection{Sphere}

The particles on a sphere experience a radial magnetic field generated by a magnetic monopole at the center. If the magnetic flux through the sphere is $N^{\sigma}_{\phi}$, the LLL single-particle wave functions are \cite{WuTT1976}
\begin{eqnarray}
\psi^{N^{\sigma}_{\phi}}_{m}(\theta,\xi) = \left[ \frac{N^{\sigma}_{\phi}+1}{4\pi} \binom{N^{\sigma}_{\phi}}{N^{\sigma}_{\phi}-m} \right]^{\frac{1}{2}} u^{N^{\sigma}_{\phi}/2+m} v^{N^{\sigma}_{\phi}/2-m}
\end{eqnarray}
where $u=\cos(\theta/2)e^{i\xi/2},v=\sin(\theta/2)e^{-i\xi/2}$ are spinor coordinates ($\theta$ and $\xi$ are the azimuthal and radial angles in the spherical coordinate system) and $m$ is the $z$ component of the angular momentum. The magnetic length is related to the radius of the sphere as $R=\ell^{+}_{B}\sqrt{N^{+}_{\phi}/2}=\ell^{-}_{B}\sqrt{N^{-}_{\phi}/2}$. The coefficients $F^{\sigma\tau\tau\sigma}_{m_{1}m_{2}m_{4}m_{3}}$ are
\begin{eqnarray}
\int d{\mathbf \Omega}_1 d{\mathbf \Omega}_2 \; \left[ \psi^{N^{\sigma}_{\phi}}_{m_1}({\mathbf \Omega}_1) \right]^* \left[ \psi^{N^{\tau}_{\phi}}_{m_2}({\mathbf \Omega}_2) \right]^* V({\mathbf r}_1-{\mathbf r}_2) \psi^{N^{\tau}_{\phi}}_{m_4}({\mathbf \Omega}_2) \psi^{N^{\sigma}_{\phi}}_{m_3}({\mathbf \Omega}_1)
\end{eqnarray}
where ${\mathbf r}=R(\sin\theta\cos\phi,\sin\theta\sin\phi,\cos\theta)$ and ${\mathbf \Omega}={\mathbf r}/R$. The product of two wave functions can be expressed as 
\begin{eqnarray}
\psi^{N^{\sigma}_{\phi}}_{m_1} \psi^{N^{\tau}_{\phi}}_{m_2} = (-1)^{N^{\sigma}_{\phi}-N^{\tau}_{\phi}} \left[ \frac{(N^{\sigma}_{\phi}+1)(N^{\tau}_{\phi}+1)}{4\pi(N^{\sigma}_{\phi}+N^{\tau}_{\phi}+1)} \right]^{1/2} \left\langle \frac{N^{\sigma}_{\phi}}{2},-m_1;\frac{N^{\tau}_{\phi}}{2},-m_2 \Bigg| \frac{N^{\sigma}_{\phi}}{2}+\frac{N^{\tau}_{\phi}}{2},-m_1-m_2 \right\rangle \psi^{N^{\sigma}_{\phi}+N^{\tau}_{\phi}}_{m_1+m_2}
\end{eqnarray}
The Coulomb potential can be expanded as
\begin{eqnarray}
\frac{e^2}{\varepsilon|{\mathbf r}_1-{\mathbf r}_2|} = \frac{4{\pi}e^2}{{\varepsilon}R} \sum^{\infty}_{L=0} \sum^{L}_{M=-L} \frac{1}{2L+1} \left[ \psi^{0}_{LM}({\mathbf \Omega}_1) \right]^* \psi^{0}_{LM}({\mathbf \Omega}_2)
\end{eqnarray}
These relations help us to obtain
\begin{eqnarray}
F^{\sigma\tau\tau\sigma}_{m_{1}m_{2}m_{4}m_{3}} = \delta_{m_1+m_2,m_3+m_4} \frac{4{\pi}e^2}{{\varepsilon}R} \sum^{{\rm min}(N^{\sigma}_{\phi},N^{\tau}_{\phi})}_{L=0} \frac{1}{2L+1} (-1)^{(N^{\sigma}_{\phi}+N^{\tau}_{\phi})/2-m_1-m_4} S^{1}_{L} S^{2}_{L}
\end{eqnarray}
where the two coefficients $S^{1,2}_{L}$ are defined by
\begin{eqnarray}
&& \left[ \psi^{N^{\sigma}_{\phi}}_{m_1}({\mathbf \Omega}_1) \right]^* \left[ \psi^{0}_{LM}({\mathbf \Omega}_1) \right]^* \psi^{N^{\sigma}_{\phi}}_{m_3}({\mathbf \Omega}_1) = \sum^{N^{\sigma}_{\phi}}_{L_1=0} (-1)^{N^{\sigma}_{\phi}/2-m_1} S^{1}_{L_1} \left[ \psi^{0}_{LM}({\mathbf \Omega}_1) \right]^* \psi^{0}_{L_1,m_3-m_1}({\mathbf \Omega}_1) \\
&& \left[ \psi^{N^{\tau}_{\phi}}_{m_2}({\mathbf \Omega}_2) \right]^* \psi^{0}_{LM}({\mathbf \Omega}_2) \psi^{N^{\tau}_{\phi}}_{m_4}({\mathbf \Omega}_2) = \sum^{N^{\tau}_{\phi}}_{L_2=0} (-1)^{N^{\tau}_{\phi}/2-m_4} S^{2}_{L_2} \left[ \psi^{0}_{L_2,m_2-m_4}({\mathbf \Omega}_2) \right]^* \psi^{0}_{LM}({\mathbf \Omega}_2)
\end{eqnarray}

\subsection{Torus}

The torus is spanned by the vectors $\mathbf{L}_{1}=L_{1}{\widehat e}_{x},\mathbf{L}_{2}=L_{2}{\widehat e}_{y}$ and we choose the Landau gauge ${\mathbf A}^{\sigma}=(0,B^{\sigma}x,0)$. If the magnetic flux through the torus is $N^{\sigma}_{\phi}$, the LLL single-particle wave functions are \cite{Yoshioka1983}
\begin{eqnarray}
\psi^{N^{\sigma}_{\phi}}_{m}(x,y) = \frac{1}{(\sqrt{{\pi}} L_2\ell^{\sigma}_{B})^{1/2}} \sum^{\mathbb{Z}}_{k} \exp \left\{ - \frac{1}{2} \left[ \frac{x}{\ell^{\sigma}_B} - \frac{2{\pi}\ell^{\sigma}_{B}}{L_2} \left( m+kN^{\sigma}_{\phi} \right) \right]^2 + i \frac{2{\pi}y}{L_2} \left( m+kN^{\sigma}_{\phi} \right) \right\}
\end{eqnarray}
where the magnetic length $\ell^{\sigma}_{B}=\sqrt{L_{1}L_{2}/(2{\pi}N^{\sigma}_{\phi})}$. By defining the reciprocal lattice vectors ${\mathbf G}_{1}=2\pi{\widehat e}_{x}/L_{1},{\mathbf G}_{2}=2\pi{\widehat e}_{y}/L_{2}$, we transform the interaction potential to momentum space as 
\begin{eqnarray}
V({\mathbf r}_1 - {\mathbf r}_2) &=& \frac{1}{L_{1}L_{2}} \sum_{\mathbf{q}} V(\mathbf q) e^{i{\mathbf q}\cdot({\mathbf r}_1 - {\mathbf r}_2)}
\end{eqnarray}
where ${\mathbf q}=q_1{\mathbf G}_1+q_2{\mathbf G}_2$. The coefficients $F^{\sigma\tau\tau\sigma}_{m_1m_2m_4m_3}$ are
\begin{eqnarray}
&& \int d^2 {\mathbf r}_1 d^2 {\mathbf r}_2 \; \left[ \psi^{N^{\sigma}_{\phi}}_{m_1}({\mathbf r}_1) \right]^* \left[ \psi^{N^{\tau}_{\phi}}_{m_2}({\mathbf r}_2) \right]^* V({\mathbf r}_1-{\mathbf r}_2) \psi^{N^{\tau}_{\phi}}_{m_4}({\mathbf r}_2) \psi^{N^{\sigma}_{\phi}}_{m_3}({\mathbf r}_1) \nonumber \\
= && \frac{1}{L_{1}L_{2}} \sum^{N^{\sigma}_{\phi}}_{m_1} \sum^{N^{\tau}_{\phi}}_{m_2} \sum_{q_1,q_2} V({\mathbf q}) \exp \left\{ -\frac{{\mathbf q}^2}{4} (\ell^{{\sigma}2}_{B}+\ell^{{\tau}2}_{B}) + i 2{\pi}q_1 \left[ \frac{(m_1-q_2/2)}{N^{\sigma}_{\phi}} - \frac{(m_2+q_2/2)}{N^{\tau}_{\phi}} \right] \right\} {\widetilde\delta}^{N^{\rm G}_{\phi}}_{m_1+m_2,m_3+m_4}
\end{eqnarray}
where $N^{\rm G}_{\phi}$ is the greatest common divisor of $N^{+}_{\phi}$ and $N^{-}_{\phi}$ and ${\widetilde\delta}^{N_{\phi}}_{i,j}$ is a generalized Kronecker delta defined as
\begin{eqnarray}
{\widetilde\delta}^{N_{\phi}}_{i,j}=1 \;\; {\rm iff} \;\; i \; {\rm mod} \; N_{\phi} = j \; {\rm mod} \; N_{\phi}
\end{eqnarray}
The many-body eigenstates are labeled by a special momentum quantum number $Y \equiv (\sum_{\sigma={\pm}}\sum^{N^{\sigma}_{\phi}}_{i=1} m^{\sigma}_{i}) \; {\rm mod} \; N^{\rm G}_{\phi}$. 

\end{widetext}

\bibliography{../ReferCollect}

\end{document}